\documentclass[conference]{IEEEtran}
\usepackage[hyphens]{url}
\usepackage{hyperref}
\hypersetup{breaklinks=true, colorlinks,allcolors=blue}
\usepackage[backend=biber, style=ieee]{biblatex}

\IEEEoverridecommandlockouts
\usepackage{amsmath,amssymb,amsfonts}
\usepackage{algorithmic}
\usepackage{graphicx}
\usepackage{textcomp}
\usepackage{xcolor}  % for colored inline text
\usepackage{balance}
\usepackage{mathtools}  % for cartesian product symbolb bigtimes
\usepackage{booktabs}  % for professional-quality tables
\usepackage{multirow}  % for multirow functionality in tables
\usepackage{orcidlink}  % for orcid in author list if wanted
\usepackage{kantlipsum} % used to generate filler text
\usepackage{float}
\usepackage{stfloats}
\usepackage{enumitem}
\usepackage{comment}

\usepackage{geometry}
\def\BibTeX{{\rm B\kern-.05em{\sc i\kern-.025em b}\kern-.08em
    T\kern-.1667em\lower.7ex\hbox{E}\kern-.125emX}}

\begin{document}

\title{An Ethically Grounded LLM-Based Approach to Insider Threat Synthesis and Detection}

\author{\IEEEauthorblockN{Haywood Gelman\,\orcidlink{0009-0009-7208-1624}, John D. Hastings\,\orcidlink{0000-0003-0871-3622}}
\IEEEauthorblockA{\textit{Beacom College of Computer \& Cyber Sciences} \\
\textit{Dakota State University }\\
Madison, SD, USA \\
haywood.gelman@trojans.dsu.edu, john.hastings@dsu.edu}

\and

\IEEEauthorblockN{David Kenley\,\orcidlink{0000-0001-7780-7644}}
\IEEEauthorblockA{\textit{College of Arts \& Sciences} \\
\textit{Dakota State University }\\
Madison, SD, USA \\
david.kenley@dsu.edu}
}

\maketitle
\begin{abstract}

Insider threats are a growing organizational problem due to the complexity of identifying their technical and behavioral elements. A large research body is dedicated to the study of insider threats from technological, psychological, and educational perspectives. However, research in this domain has been generally dependent on datasets that are static and limited access which restricts the development of adaptive detection models. This study introduces a novel, ethically grounded approach that uses the large language model (LLM) Claude Sonnet 3.7 to dynamically synthesize syslog messages, some of which contain indicators of insider threat scenarios. The messages reflect real-world data distributions by being highly imbalanced (1\% insider threats). The syslogs were analyzed for insider threats by both Sonnet 3.7 and GPT-4o, with their performance evaluated through statistical metrics including accuracy, precision, recall, F1, specificity, FAR, MCC, and ROC AUC.  Sonnet 3.7 consistently outperformed GPT-4o across nearly all metrics, particularly in reducing false alarms and improving detection accuracy. The results show strong promise for the use of LLMs in synthetic dataset generation and insider threat detection.
\end{abstract}


\section{Introduction}

Malicious insider threats are a human problem that manifest in computer networks as data loss, theft, or destruction \cite{schoenherr_multiple_2022}. Motivations for malicious insider threat activities include financial gain \cite{nurse_understanding_2014}, retribution for perceived fault \cite{greitzer_identifying_2012}, and rationalization \cite{padayachee_joint_2021}, sometimes driven by subclinical traits \cite{harms_exposing_2022}. The evasive nature and small number of insider threats creates detection difficulties for organizations \cite{greitzer_insider_2019}. Insider threats are minimally observed on corporate networks, with centralized system logging (syslog) as an important aggregation tool to enable discovery \cite{yamanishi_dynamic_2005}. Organizations rely on many detection solutions in addition to syslog, including security incident event monitoring (SIEM) \cite{vaisanen_categorization_2017}, user entity behavioral analysis (UEBA) \cite{alshehhi_scenario_2023}, machine learning-based log analysis \cite{alzaabi_review_2024}, ML-driven log inspection \cite{younesian_syslog_2021}, and human intervention \cite{gelman_toward_2024}. 

An emerging area of insider threat research is the use of large language models (LLMs) to detect insider threats in existing datasets \cite{gelman_scalable_2025}. However, access to realistic datasets for research purposes can be limited due to the proprietary, and confidential nature of such data. Further, the human element of insider threat research requires careful attention to data privacy concerns which might prevent the release of real-world data.

This research addresses the ethical and access issues of real-world data by leveraging LLMs to generate a unique synthetic syslog dataset with dynamically created log content. Synthetic message generation with commonly used syslog fields for validity and LLM-generated code elements ensures a novel, realistic approach while maintaining data privacy. Insider threat detection tests are then performed by LLMs on the data.
The study includes an imbalanced mix of standard (non-insider threat) and insider threat syslog messages to mimic, with minor limitations, realistic syslog messages for LLM inspection. To evaluate LLM detection accuracy within an acceptable margin of error, a small percentage of the log population (1\%) was dynamically generated as insider threat logs.

According to an exhaustive literature search, this study appears to be the first of its kind to use the methodology %\textbf{and experimental design} 
described in this paper. The primary novel contributions of this insider threat research are: 
\begin{enumerate}
    \item addressing limited access to real-world data by utilizing LLMs to synthesize syslog messages for insider threat indicators, and
    \item producing a more comprehensive analysis of LLM-automated detection of insider threats than previously seen.
\end{enumerate}
The following research questions guide the study: 
\begin{itemize}
    \item \textbf{RQ1}: How effective are LLMs in identifying insider threats within synthetically generated syslog messages?
    \item \textbf{RQ2}: What is the comparative performance between the LLMs for highly imbalanced syslog datasets?
    \item \textbf{RQ3}: What are the comparative false alarm rates when utilizing LLM-based insider threat detection? 
\end{itemize}

\section{Methodology}
\label{meth}

The methodology follows a three-phase process: 1) generation of synthetic syslog datasets, 2) analysis of the logs for insider threats by the LLMs via API, and 3) statistical analysis of detection results. Each phase is described below.

\subsection{Synthetic Syslog Generation}

Syslog messages were generated through a menu-driven program (SysGen)\footnote{LLMs generated the preliminary Python code for this research with direct human interaction, inspection, debugging, and rigorous validity testing.} to configure the number of standard logs, the number of insider threat logs, the syslog server IP and port, and export filenames for raw and structured output. Cochran's sample size formula \cite{cochran1977sampling} applied to a population of 10 million messages dictated a sample size of 385. Two datasets were generated: one control dataset with 385 standard logs and no insider threat messages, and one 385 syslog message dataset with 381 standard logs (negative class) and four insider threat logs (1\%, positive class). These imbalanced datasets are intended to reflect the real-world rarity of insider threats and formed the basis for this research. 

The syslog server required SIEM compatible syslog messages with basic formatting logic and an embedded JSON-formatted message field. 
Existing datasets such as CERT \cite{carnegie_mellon_university_insider_2020} are widely used in insider threat research but were not appropriate for this study due to field limitations and lack of SIEM-level integration. The logs include realistic field values drawn from RFC 3164 and 5424 standards, with validation references from Palo Alto Networks \cite{palo_alto_networks_inc_pan-os_2025}, CISA \cite{cisa_cisa_2024}, the MITRE ATT\&CK® Framework \cite{mitre_attck_mitre_2024}, and NIST 800-53r5 \cite{joint_task_force_interagency_working_group_security_2020}. Fields include: timestamp, username, session id, auth method, src ip, src hostname, action, object, resource, command, status, bytes, app name, dst ip, dst port, protocol, duration, network zone, location category, criticality, is approved application, and previous occurrence count.

Each log entry includes a unique session ID for validation, although session ID is not part of the standard syslog format. Logs were exported in structured CSV format with optional raw text output for inspection. Integrity checking code analyzed the unique dataset session IDs for loss, duplication, and malformed data, while visual inspection mitigated validity concerns. 

\subsection{LLM Insider Threat Syslog Analysis}
\label{sec:log-analysis}
Sonnet 3.7 \cite{anthropic_claude_2025} and %OpenAI's 
GPT-4o \cite{GPT4o} were selected as the LLMs to analyze the synthetic logs given their availability, their API integration\footnote{Sonnet 3.7 and GPT-4o generated code to interface with their respective APIs.}, cost, and prior strong performance in the related task of sentiment analysis \cite{gelman_scalable_2025, hastings_utilizing_2024, weitl-harms_using_2024}. Log analysis consisted of acquiring the output dataset from syslog generation, validating the contents, and executing code to send the dataset to the LLM through API for scoring. LLMs returned a scored dataset with fields including session\_id, true\_label (empty), and predicted\_label. The dataset also contained non-syslog fields confidence\_score and classification\_result for use in future research. These two fields were not utilized in this research and were not included in log predictions or statistical analysis. The control and intervention datasets were sent to the LLMs, once each for the control dataset and three times each for the intervention dataset. Large log files were split as needed to fit within the available context window. Post-scoring, the datasets were aggregated for statistical analysis. 

\subsection{Statistical Evaluation}
\label{sec:stats-analysis}

Statistical analyses evaluated model accuracy, recall, precision, false alarm rate (FAR), Matthews Correlation Coefficient (MCC) \cite{chicco_advantages_2020}, F1, specificity, and ROC AUC \cite{fawcett_introduction_2006}. Accuracy and recall determine overall performance and detection capability, while FAR and MCC provide insight into model behavior in high-class imbalance.

\section{Experimental Design}
\label{sec:experimental-design}

\begin{figure*}[htbp]
\centering
\includegraphics[width=0.70\linewidth]{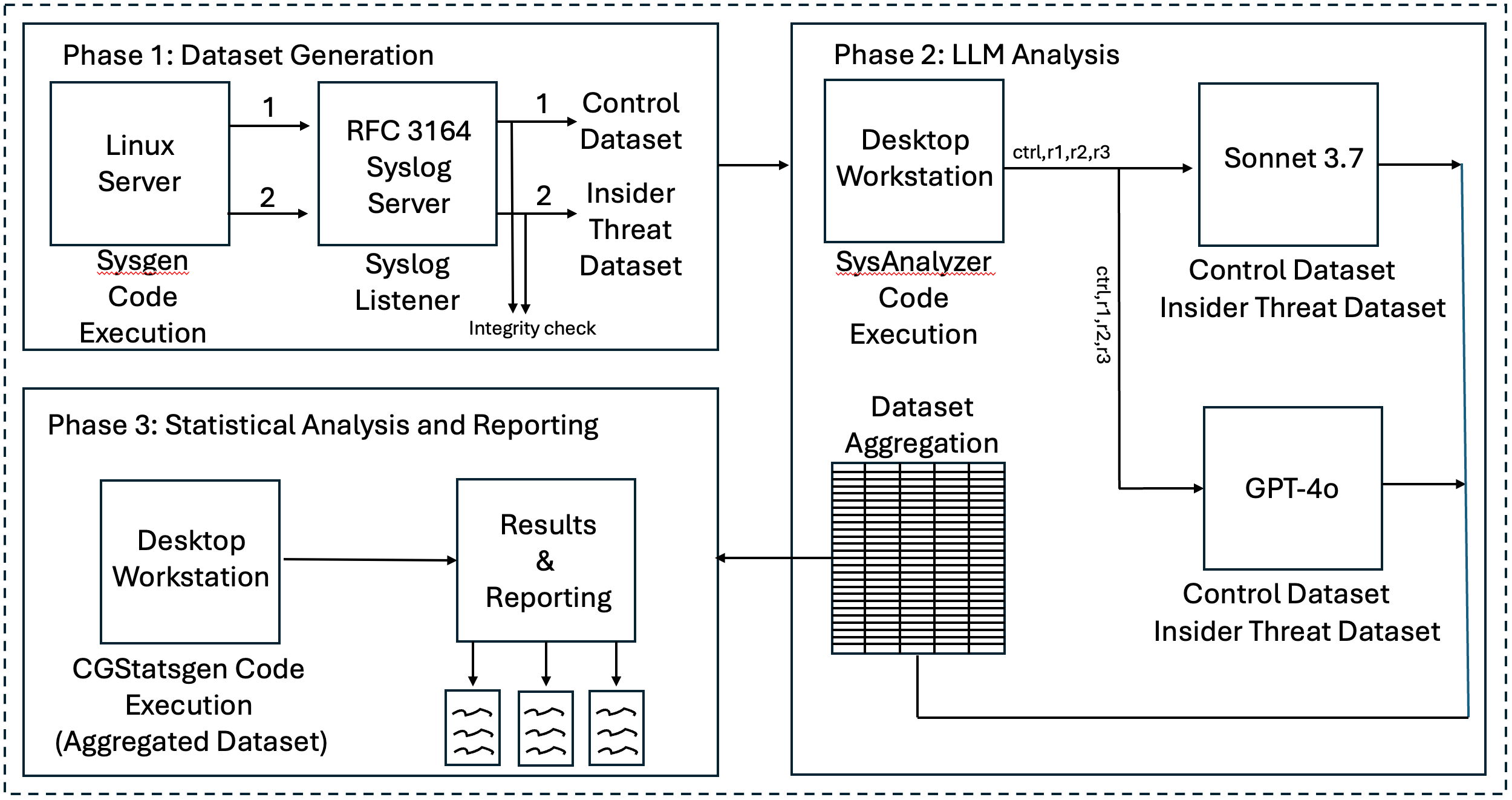}
\caption{Insider Threat Syslog Research Design}
\label{fig:exp-design-phases.png}
\end{figure*}

The experimental design consists of three phases (Fig. \ref{fig:exp-design-phases.png}): dataset generation, LLM analysis, and statistical analysis \& reporting. In Phase 1, the operator provides SysGen the IP address and port of the syslog server, how many logs to send, and where to save CSV and raw log output files. The code sends the specified logs to the syslog server, where the logs are collected, exported, and visually validated against the output CSV logs. Phase 1 is executed twice, once to generate a control dataset with standard (non-insider threat) logs and once to generate a specified combination of standard and insider threat logs. Once log transmission completes, log integrity code validates log files based on the unique session\_id, is visually inspected for good data, and prepared for Phase 2. 

In Phase 2, a blank true\_label column is inserted in the control and treatment datasets with instructions to the LLMs to copy it to the output file and ignore it. SysAnalyzer executes separately with each dataset, and asks the operator for an API key, input CSV file name, and output CSV file name. Python code sends the control dataset and three treatment dataset runs to each LLM, and saves output CSVs for visual verification and validity. Output CSV columns include session\_id, true\_value (blank), predicted\_value, and confidence\_score. Fields from all eight datasets are aggregated into a single dataset for statistical analysis. Aggregated column headers include control\allowbreak\_session\allowbreak\_id, control\allowbreak\_true\allowbreak\_label, treatment\allowbreak\_session\allowbreak\_id, treatment\allowbreak\_true\allowbreak\_label, claude\allowbreak\_predicted\allowbreak\_control, claude\allowbreak\_predicted\allowbreak\_int\allowbreak\_run1, claude\allowbreak\_predicted\allowbreak\_int\allowbreak\_run2, claude\allowbreak\_predicted\allowbreak\_int\allowbreak\_run3, gpt4o\allowbreak\_predicted\allowbreak\_control, gpt4o\allowbreak\_predicted\allowbreak\_int\allowbreak\_run1, gpt4o\allowbreak\_predicted\allowbreak\_int\allowbreak\_run2, and gpt4o\allowbreak\_predicted\allowbreak\_int\allowbreak\_run3.

In Phase 3, CGStatsGen processes the aggregated dataset to generate results for statistical analyses noted in \ref{sec:stats-analysis} in the form of LaTeX code, CSV, and figures. 

\section{Experimental Configuration}
\label{exp-cfg}
The experimental configuration includes the workstation and syslog server network configuration, Ubuntu server Python and networking configuration, Python code configuration on the desktop workstation for API calls, and Python code configuration for statistical analysis.

\subsection{SysGen}

SysGen code is executed on a Linux server running Ubuntu 24.04 sending syslog messages to a SyslogView syslog server version 2.1.1 with UDP port 514 on macOS 14.7.1. The server sends syslog messages from IP address 192.168.1.10/24 to 192.168.1.20/24. The subnet is isolated with protected internet access. The server runs Python 3.12.3 in a virtual environment with default packages socket, random, time, datetime, json, uuid, argparse, sys, csv, os, and IP address. Syslog messages are exported from SyslogView and copied from server to workstation with a cloud storage solution. {Fig. \ref{fig:ec-phase1-figure1.png} depicts the Phase 1 experimental configuration. 

\begin{figure}[htbp]
\centering
\includegraphics[width=0.65\columnwidth]{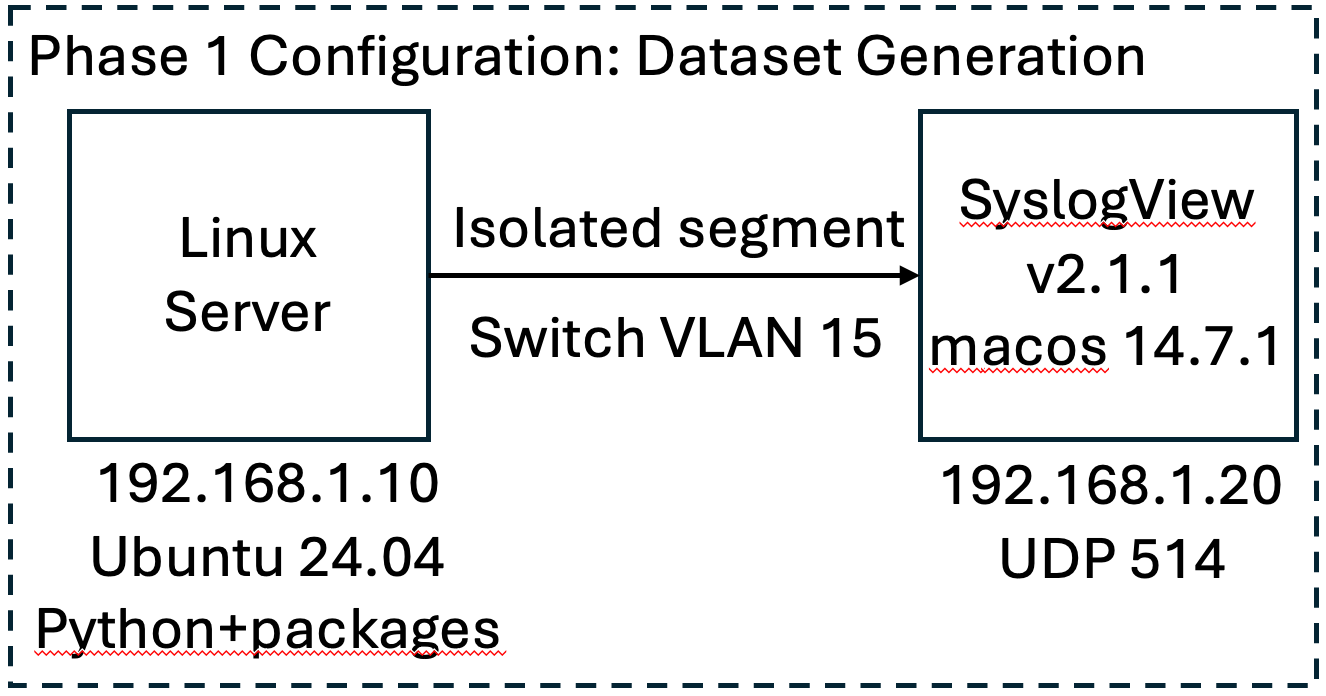} 
\caption{Insider Threat Syslog Research Design, Phase 1}
\label{fig:ec-phase1-figure1.png}
\end{figure}

\subsection{SysAnalyzer}

SysAnalyzer code is executed on a macOS 14.7.1 workstation. The workstation runs Python 3.9.6 and uses packages requests, pandas, time, datetime, json, warnings, stringio, numpy, os, logging, and traceback for API access to GPT-4o. For access to Sonnet 3.7's API, Python 3.9.6 imports packages os, sys, json, requests, anthropic, re, csv, time, random, pandas, math, and path. For each API, the respective code is executed and asks for the API key, the input CSV file, and the output file name. {Fig. \ref{fig:ec-phase2-figure2.png} describes the Phase 2 experimental design. 

\begin{figure}[htbp]
\centering
\includegraphics[width=0.65\columnwidth]{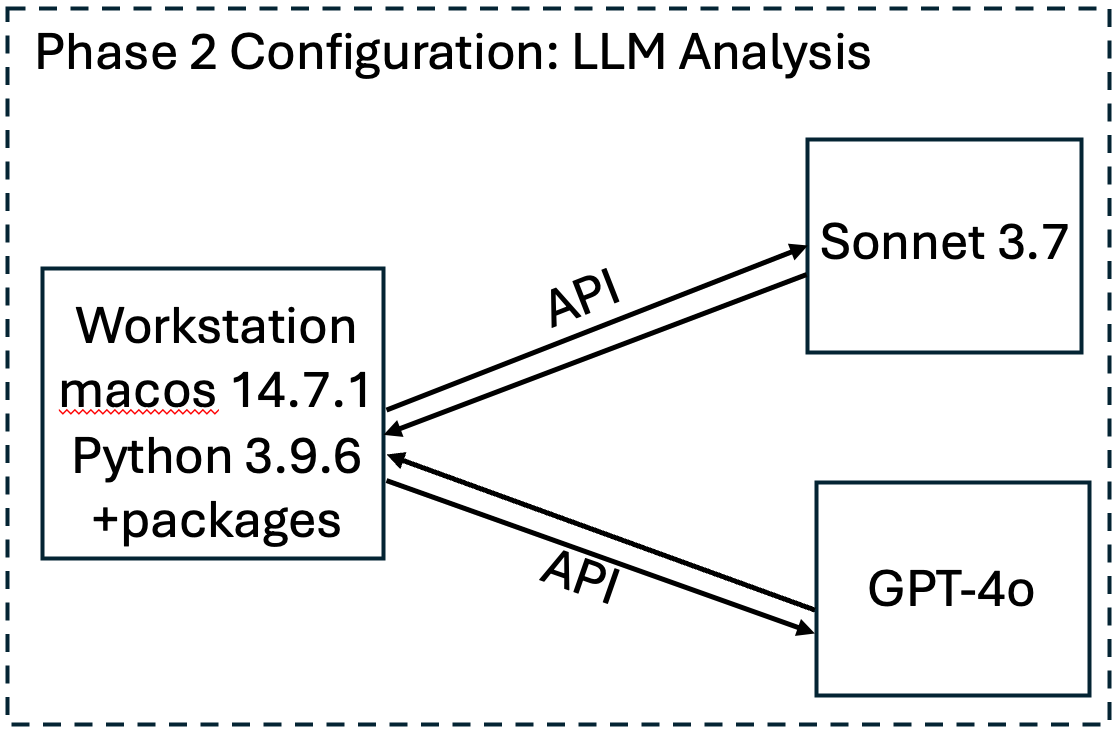}
\caption{Insider Threat Syslog Research Design, Phase 2}
\label{fig:ec-phase2-figure2.png}
\end{figure}

\subsection{CGStatsGen}

At CGStatsGen code execution, Python loads packages os, pandas, numpy, matplotlib.pyplot, seaborn, proportion\allowbreak\_confint, binom\_test, accuracy\allowbreak\_score, precision\allowbreak\_score, recall\allowbreak\_score, f1\allowbreak\_score, confusion\allowbreak\_matrix, matthews\allowbreak\_corrcoef, precision\allowbreak\_recall\allowbreak\_curve, average\allowbreak\_precision\allowbreak\_score, roc\allowbreak\_curve, auc,
datetime, and traceback. Code runs the above test modules on the binary, imbalanced dataset and exports a series of results. Results include confusion matrices, ROC curves for control and treatment runs 1-3, and a comprehensive metrics summary. The metrics summary includes model run, accuracy, precision, recall, F1 score, specificity, MCC, FAR and AUC for GPT-4o and Sonnet 3.7's control and treatment runs 1-3. {Fig. \ref{fig:ec-phase3-figure3.png} describes the Phase 3 experimental configuration. 

\begin{figure}[htbp]
\centering
\includegraphics[width=0.70\columnwidth]{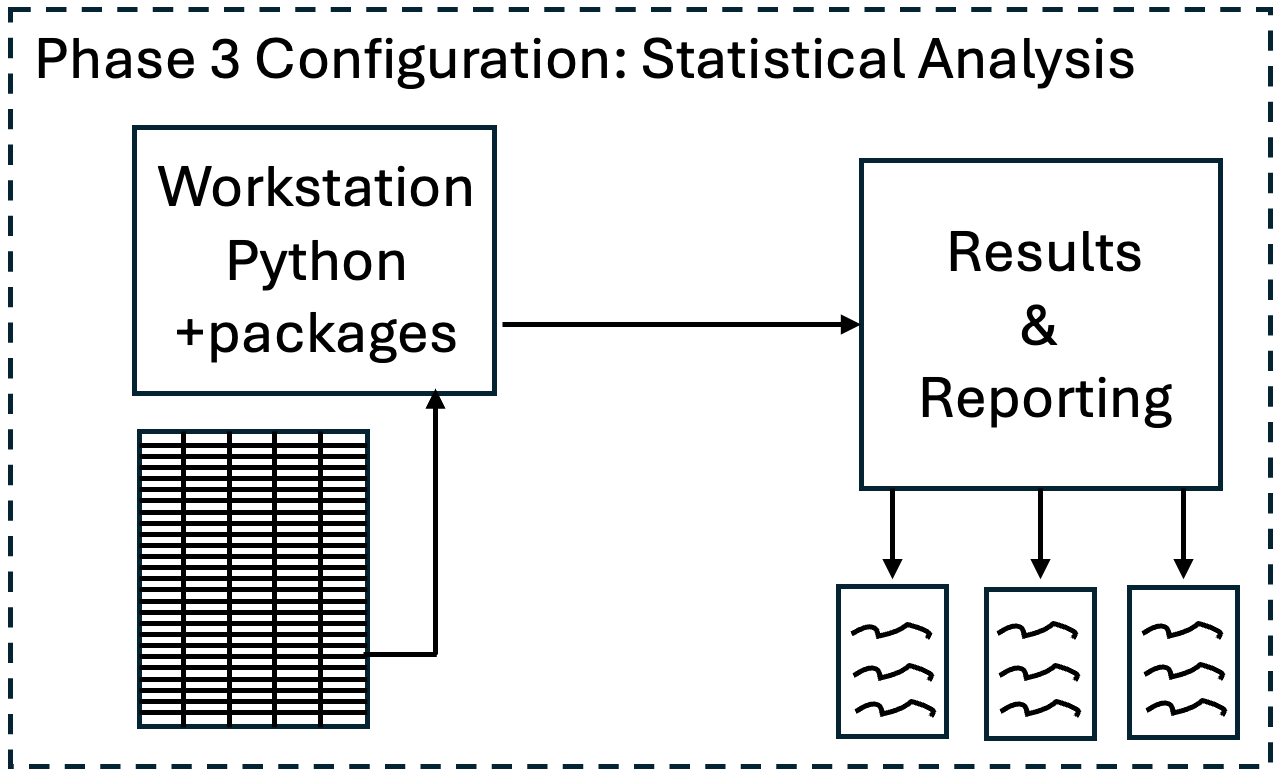}
\caption{Insider Threat Syslog Research Design, Phase 3}
\label{fig:ec-phase3-figure3.png}
\end{figure}

\section{Results}
\label{sec:results}

False positives (FP) and false negatives (FN) play an important role in evaluating model performance, especially given the class imbalance in the dataset. In answering RQ1, as shown in Table \ref{tab:confusion-matrix-summary}, both models successfully detected all insider threats (0 FN). Table \ref{tab:metrics-summary} demonstrates considerably higher false positives for GPT-4o than Sonnet 3.7, contributing to significantly higher FAR for GPT-4o than Sonnet 3.7.

\begin{table}[htbp]
\centering
\caption{Confusion matrix results showing TP, FN, FP, TN for each model}
\label{tab:confusion-matrix-summary}
\begin{tabular}{lcccccc}
\toprule
Model & TP & FN & FP & TN & Total Pos/Neg \\
\midrule
Sonnet 3.7 Control Run & 0 & 0 & 45 & 340 & 0/385 \\
Sonnet 3.7 Int Run 1 & 4 & 0 & 39 & 342 & 4/381 \\
Sonnet 3.7 Int Run 2 & 4 & 0 & 35 & 346 & 4/381 \\
Sonnet 3.7 Int Run 3 & 4 & 0 & 42 & 339 & 4/381 \\
GPT-4o Control Run & 0 & 0 & 177 & 208 & 0/385 \\
GPT-4o Int Run 1 & 4 & 0 & 169 & 212 & 4/381 \\
GPT-4o Int Run 2 & 4 & 0 & 165 & 216 & 4/381 \\
GPT-4o Int Run 3 & 4 & 0 & 174 & 207 & 4/381 \\
\bottomrule
\end{tabular}
\end{table}

\begin{table*}[htbp]
\centering
\caption{Summary of classification metrics for all model runs}
\label{tab:metrics-summary}
\begin{tabular}{lccccccc}
\toprule
Model & Accuracy & Precision & Recall & F1 & Specificity & FAR & MCC \\
\midrule
Sonnet 3.7 Control & 0.883 & NaN & NaN & NaN & 0.883 & 0.117 & NaN \\
Sonnet 3.7 Int Run 1 & 0.899 & 0.093 & 1.000 & 0.170 & 0.898 & 0.102 & 0.289 \\
Sonnet 3.7 Int Run 2 & 0.909 & 0.103 & 1.000 & 0.186 & 0.908 & 0.092 & 0.305 \\
Sonnet 3.7 Int Run 3 & 0.891 & 0.087 & 1.000 & 0.160 & 0.890 & 0.110 & 0.278 \\
GPT-4o Control & 0.540 & NaN & NaN & NaN & 0.540 & 0.460 & NaN \\
GPT-4o Int Run 1 & 0.561 & 0.023 & 1.000 & 0.045 & 0.556 & 0.444 & 0.113 \\
GPT-4o Int Run 2 & 0.571 & 0.024 & 1.000 & 0.046 & 0.567 & 0.433 & 0.116 \\
GPT-4o Int Run 3 & 0.548 & 0.022 & 1.000 & 0.044 & 0.543 & 0.457 & 0.110 \\
\bottomrule
\end{tabular}
\end{table*}

\subsection{Accuracy, Precision, and False Alarm Rate}

Table \ref{tab:metrics-summary} summarizes key classification metrics across all runs, including accuracy (95\% confidence interval), precision, recall, F1 score, specificity, MCC, FAR, and AUC, for both Sonnet 3.7 and GPT-4o. Accuracy is a proportional measure of total correct true positive (TP) and true negative (TN) predictions with all predictions as the divisor \cite{google_developers_classification_nodate}. Table \ref{tab:model-comparison} shows a pairwise comparison of the accuracy of the two LLMs for each run. In answering RQ2, Sonnet 3.7 consistently outperforms GPT-4o in all runs with  accuracy improvements of \(\sim\!34\)\%.
Sonnet 3.7's average intervention accuracy was 0.899, and GPT-4o's average was 0.560. Note, however, that with an intervention dataset of 381 standard logs and four insider threat logs, a nearly perfect accuracy score is achievable for an LLM by simply predicting only TN values. 
Accuracy alone can thus be misleading when applied to imbalanced binary datasets.

Dataset imbalance does not invalidate accuracy, but it requires the use of other measures to add additional perspective. Precision, a TP prediction measure when an insider threat is present, is considered in tandem with FAR \cite{sokolova_systematic_2009}. Together, these measurements determine how often models predict a threat when no threat is present, but from opposite perspectives \cite{powers_evaluation_2020}. Organizations use precision and FAR  to measure when a system falsely records a threat, which increases the number of alerts and reduces system reliability \cite{google_developers_classification_nodate}. Sonnet 3.7's average precision to FAR measure for intervention runs was 0.094/0.101. This indicates that, with this dataset, the model underperforms in predictions with slightly higher than 10\% false alarms. This compares to GPT-4o, whose precision/FAR average for intervention runs was 0.023/0.445, demonstrating lower precision and a higher FAR than Sonnet 3.7.

\subsection{Specificity, False Alarm Rate, Recall, and F1}
Similar to accuracy and precision, specificity and FAR are considered in tandem as a measure of TN predictions. Specificity and FAR measure the correct identification of non-threats on a 0.0-1.0 scale (including control runs) \cite{powers_evaluation_2020}. In answering RQ3, Sonnet 3.7's specificity/FAR average was 0.899/0.101, indicating high performance on correctly identifying TN with a 10\% error rate. GPT-4o's average specificity/FAR score was 0.555/0.445, demonstrating overly conservative precision behavior coupled with a very high FAR. This measure is in direct contrast with recall, a measure of prediction on TP, which shows perfect prediction on the four TP threats for both models. F1 score accounts for both positive predictions and TP by balancing precision and recall through harmonic mean calculation \cite{chicco_advantages_2020}. As a result of the harmonic mean calculation, low precision, and high recall, average F1 scores are low for Sonnet 3.7 and GPT-4o, respectively, at 0.172 and 0.045. Research determined that an imbalanced dataset proved problematic in these areas with a dataset containing an intentionally small number of TP.

\subsection{Matthews Correlational Coefficient}
MCC generates inclusive statistical analyses to support imbalanced datasets. Unlike previous statistics, MCC includes all predictions as part of its calculations on a [-1,1] scale, with scores above 0.5 demonstrating reasonable reliability. Sonnet 3.7's average MCC score at 0.291 demonstrates reliability below the minimum threshold of 0.3 as marginally acceptable, although run2 was 0.305. GPT-4o scored far lower at 0.113. 

\subsection{ROC AUC}
Like MCC, ROC AUC incorporates both TP and TN predictions. The ROC curve plots the TP rate against the FP rate at various thresholds, and AUC quantifies the overall ability of the model to distinguish between classes. An AUC of 0.5 indicates random guessing, while values closer to 1.0 indicate stronger performance. ROC's effective threshold is 0.0-1.0, and AUC is 0.5-1.0, with 0.5 representing a random occurrence. For ROC, a per-prediction score is computed as $s=p$ if $y=1$ else $1-p$, where $y$ is the true label and $p\in[0,1]$ is the model confidence. Fig. \ref{fig:roc_curves_run1.png} depicts the performance difference between the two models with improved indications toward the top left corner. The blue line represents Sonnet 3.7 performance, and the orange line represents GPT-4o.  At 0.994, Sonnet 3.7's average AUC score showed strong performance with an improvement of \(\sim\!3.5\)\% over GPT-4o. GPT-4o, although not as strong, still demonstrated good performance according to this measure at 0.960. 

\begin{figure}[htbp]
\centering
\includegraphics[scale=.5]{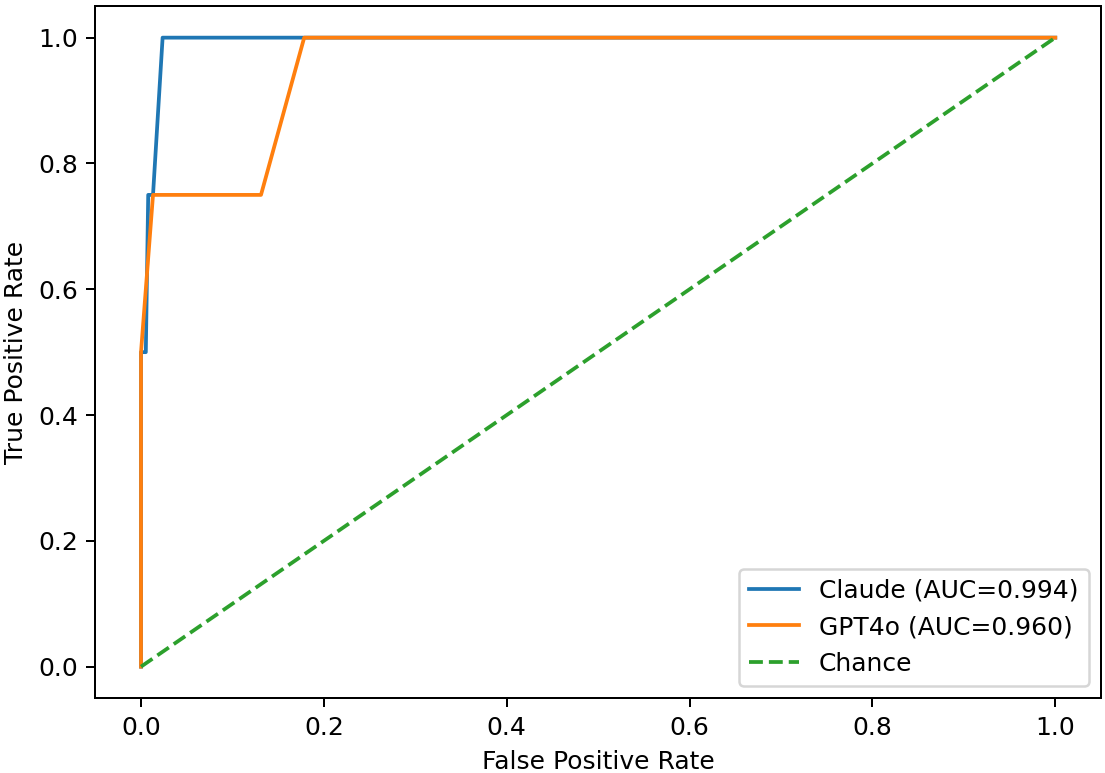}
\vspace{-0.8em}
\caption{ROC Curve Comparison}
\label{fig:roc_curves_run1.png}
\end{figure}

\begin{comment}
\subsection{Fisher's Exact Test and Cohen's H}

Fisher's exact test is used to calculate the probability that models generated predictions by random chance, accounting for all predictions in a dataset \cite{bower_when_2020}. Fisher's test is the ideal test to calculate p-value with small, binary, imbalanced datasets due to its consideration of all predicted values \cite{bower_when_2020}. Table \ref{tab:model-comparison}'s statistical analysis demonstrates Fisher's exact test application of Table \ref{tab:confusion-matrix-summary}'s contingency table summary.  Analysis generates a p-value to enable a performance comparison between Sonnet 3.7 and GPT-4o . The calculated p-values are stated in scientific notation to compactly and accurately represent the small p-values. The p-value provides vigorous evidence to reject the null hypothesis.
\end{comment}

\begin{comment}
\begin{table*}[htbp]
\centering
\caption{Comparison of model accuracy, statistical significance, and effect size across runs.}
\label{tab:model-comparison}
\begin{tabular}{l*{7}{c}}
\toprule
run & Sonnet 3.7 accuracy & GPT4-o Accuracy & Accuracy diff & Fisher p-value & Odds Ratio & Cohens h & Effect \\
\midrule
control & 0.8831 & 0.5403 & 0.3429 & 1.218566e-26 & 6.43 & 0.7924 & Medium \\
int run1 & 0.8987 & 0.5610 & 0.3377 & 4.857538e-27 & 6.94 & 0.8006 & Large \\
int run2 & 0.9091 & 0.5714 & 0.3377 & 7.693595e-28 & 7.50 & 0.8149 & Large \\
int run3 & 0.8909 & 0.5481 & 0.3429 & 3.810396e-27 & 6.73 & 0.8013 & Large \\
\bottomrule
\end{tabular}
\end{table*}
\end{comment}

\begin{table}[htbp]
\centering
\caption{Comparison of model accuracy and difference across runs.} 
\label{tab:model-comparison}
\begin{tabular}{l*{7}{c}}
\toprule
Run & Sonnet 3.7 Accuracy & GPT-4o Accuracy & Accuracy Diff \\ 
\midrule
control & 0.883 & 0.540 & 0.343 \\ 
int run1 & 0.899 & 0.561 & 0.338 \\ 
int run2 & 0.909 & 0.571 & 0.338 \\ 
int run3 & 0.891 & 0.548 & 0.343 \\ 
\bottomrule
\end{tabular}
\end{table}

\section{Discussion}
\label{discussion}
The significance of this research is that it advances the field of insider threat detection by demonstrating the feasibility of using LLMs to generate and analyze synthetic syslog data in a controlled, ethical pipeline. Across all tests, Sonnet 3.7 outperformed GPT-4o in nearly every metric, with particularly notable differences in FAR with Sonnet lower by a factor of four. While formal statistical significance testing is reserved for the forthcoming journal version, results were consistent across all runs and suggest meaningful performance differences between the models. Some traditional metrics such as accuracy and F1 score proved misleading due to dataset imbalance. MCC demonstrated marginal reliability, while Sonnet 3.7 showed improved FAR performance over GPT-4o by a factor greater than 4X.

\subsection{Insider Threat Logs}
This study revealed LLMs’ ability to detect subtle distinctions between benign and potentially harmful behavior. Table \ref{control-raw-log} is an example of a standard log in raw CSV format from the control dataset which shows activity that might be viewed as an insider threat (certificate auth, object access failure on permissions change for a  web server config file). However, both LLMs and the control set registered this as TN from the standard\_operation field with a sanctioned application. 

\begin{table}[h!]
\caption{Control Dataset Raw Log}
\label{control-raw-log}
\centering
\begin{tabular}{|p{8.4cm}|}
\hline
Apr 01 19:37:41, apache, c43c337e-7de9-4e6a-aace-325c5a004ea6, certificate, 192.168.252.72, thin-client01, access, /var/www/html/index.php, file-system, chmod 644 /etc/config.xml, failure, 38775, named, 172.16.229.198, 58449,SFTP, 716, internal, office, low, True, 19
\\ \hline
\end{tabular}
\end{table}

Table \ref{int-raw-log} is a TP log identified by both LLMs as insider threat activity. This insider threat log demonstrates an attempt to establish a suspicious shell with an unsanctioned application. However, the syntax is incorrect, hence the ``False'' marker in the log. This log is an example of an attack in progress that failed due to a syntax error.

\begin{table}[h!]
\caption{Insider Threat Dataset Raw Log}
\label{int-raw-log}
\centering
\begin{tabular}{|p{8.4cm}|}
\hline
Apr 01 19:47:59, rbrown, a230d494-1647-4e50-ab74-7605d20439ac, kerberos, 172.16.41.101, tablet01, access,system,file-system, "perl -e 'exec ""/bin/sh"";'", success, 2491, oracle,10.0.138.167, 7523, SFTP, 1732, internal, office, high, False, 4
\\ \hline
\end{tabular}
\end{table}

\subsection{RFC Compliance and Forged Timestamps}
The synthetic logs follow RFC 3164 formatting, omitting fields like proc\_id and priority required in RFC 5424. In addition, logs were not organized into flow-specific sequences across IPs, ports, or timestamps, limiting multi-log incident context. Although forging timestamps is possible in RFC-based logs, and could prove useful for insider threat research, it was not used here, as API constraints required processing in 32 row batches, making such context minimally useful. Despite these limitations, the dataset structure remained valid for proper analysis of individual log entries.

\section{Future Work}
\label{sec:future}
The results demonstrated better predictive performance by Sonnet 3.7 versus GPT-4o for this methodology. The research pipeline noted in Section \ref{sec:experimental-design} described an issue with the true\_label placed in the dataset prior to LLM analysis. This early pipeline choice invalidated the data and required a prompt change and retesting to correct it. In addition, several statistics generated by the research demonstrated misleading results when observed independently. Future work will insert the true\_label after the analysis is complete and modify the prompt appropriately. Also, in future insider threat research with binary imbalanced datasets, primary statistical analysis will be Fisher's exact test, effect size, Cohen's h, MCC, PR-AUC with confidence scores, and FAR with the remaining tests utilized as secondary checks for validity.

\section{Related Work}
\label{sec:related-work}
Prior work on synthetic log generation \cite{cruikshank_cruikshank25security-log-generator_2025,forge_swimlanesoc-faker_2025} produced logs using static templates or event simulation. These generators lack the ability to dynamically create contextual log messages tailored to specific threat scenarios. Recent studies explore the use of LLMs for log parsing and anomaly detection \cite{le_log_2023,liu_interpretable_2024}, and template-based detection \cite{vaarandi_using_2024}, but do not focus on LLM-driven generation of syslog data. This research addresses that gap by using LLMs to generate semantically varied, SIEM-compatible log messages. In addition, the use of real log datasets raises challenges around ethical use, limited field detail, and lack of contextual variety \cite{glasser_bridging_2013}. By generating synthetic logs, this research supports ethical, privacy-preserving experimentation in insider threat detection.

\section{Conclusion}
\label{conclusion}
Insider threats in syslog messages are a growing research trend with promising approaches in detection methods, theories, and themes. This research introduced a novel, comprehensive method for insider threat log generation, syslog analysis, and statistical evaluation using LLMs as both data synthesizers and detection agents. The ethical research methodology used simulated logs, with fundamental RFC feature adherence in a way that is repeatable, adaptable, and robust when combined with the detailed experimental design. Results demonstrated the feasibility of using LLMs for generation, detection, and analysis of realistic, emulated insider threat syslog messages.

\printbibliography

\end{document}